\documentclass[dvips]{acta}
\usepackage{supertabular,lscape,epsfig}
\usepackage{amssymb}
\usepackage{amsmath}
\usepackage{wasysym}
\DeclareSymbolFont{ppa}{OT1}{ppl}{m}{it}
\DeclareMathSymbol{\vv}{\mathalpha}{ppa}{'166}

\newfont{\hb}{rphvb at 10pt}
\newfont{\hbo}{rphvbo at 10pt}
\newfont{\bitt}{rptmbi at 12pt}
\newfont{\bits}{rptmbi at 11pt}

\SetPages{335}{347}

\SetVol{59}{2009}

\begin{document}

\newcommand{\TabCapp}[2]{\begin{center}\parbox[t]{#1}{\centerline{
  \small {\spaceskip 2pt plus 1pt minus 1pt T a b l e}
  \refstepcounter{table}\thetable}
  \vskip2mm
  \centerline{\footnotesize #2}}
  \vskip3mm
\end{center}}

\newcommand{\TTabCap}[3]{\begin{center}\parbox[t]{#1}{\centerline{
  \small {\spaceskip 2pt plus 1pt minus 1pt T a b l e}
  \refstepcounter{table}\thetable}
  \vskip2mm
  \centerline{\footnotesize #2}
  \centerline{\footnotesize #3}}
  \vskip1mm
\end{center}}

\newcommand{\MakeTableSepp}[4]{\begin{table}[p]\TabCapp{#2}{#3}
  \begin{center} \TableFont \begin{tabular}{#1} #4 
  \end{tabular}\end{center}\end{table}}

\newcommand{\MakeTableee}[4]{\begin{table}[htb]\TabCapp{#2}{#3}
  \begin{center} \TableFont \begin{tabular}{#1} #4
  \end{tabular}\end{center}\end{table}}

\newcommand{\MakeTablee}[5]{\begin{table}[htb]\TTabCap{#2}{#3}{#4}
  \begin{center} \TableFont \begin{tabular}{#1} #5 
  \end{tabular}\end{center}\end{table}}

\newfont{\bb}{ptmbi8t at 12pt}
\newfont{\bbb}{cmbxti10}
\newfont{\bbbb}{cmbxti10 at 9pt}
\newcommand{\uprule}{\rule{0pt}{2.5ex}}
\newcommand{\douprule}{\rule[-2ex]{0pt}{4.5ex}}
\newcommand{\dorule}{\rule[-2ex]{0pt}{2ex}}
\def\thefootnote{\fnsymbol{footnote}}
\begin{Titlepage}
\Title{The Optical Gravitational Lensing Experiment.\\
The OGLE-III Catalog of Variable Stars.\\
V.~~R Coronae Borealis Stars in the Large Magellanic Cloud\footnote{Based on
observations obtained with the 1.3-m Warsaw telescope at the Las Campanas
Observatory of the Carnegie Institution of Washington.}}
\Author{I.~~S~o~s~z~y~ñ~s~k~i$^1$,~~
A.~~U~d~a~l~s~k~i$^1$,~~
M.\,K.~~S~z~y~m~a~ñ~s~k~i$^1$,\\
M.~~K~u~b~i~a~k$^1$,~~
G.~~P~i~e~t~r~z~y~ñ~s~k~i$^{1,2}$,~~
£.~~W~y~r~z~y~k~o~w~s~k~i$^3$,\\
O.~~S~z~e~w~c~z~y~k$^2$,
~~K.~~U~l~a~c~z~y~k$^1$~~
and~~R.~~P~o~l~e~s~k~i$^1$}
{$^1$Warsaw University Observatory, Al.~Ujazdowskie~4, 00-478~Warszawa, Poland\\
e-mail:
(soszynsk,udalski,msz,mk,pietrzyn,wyrzykow,kulaczyk,rpoleski)
@astrouw.edu.pl\\
$^2$ Universidad de Concepci{\'o}n, Departamento de Fisica, Casilla 160--C,
Concepci{\'o}n, Chile\\
e-mail: szewczyk@astro-udec.cl\\
$^3$ Institute of Astronomy, University of
Cambridge, Madingley Road, Cambridge CB3 0HA, UK\\
e-mail: wyrzykow@ast.cam.ac.uk}
\Received{November 20, 2009}
\end{Titlepage}
\Abstract{The fifth part of the OGLE-III Catalog of Variable Stars presents
23 R~CrB (RCB) stars in the Large Magellanic Cloud (LMC). 17 of these
objects have been spectroscopically confirmed by previous studies, while 6
stars are new candidates for RCB variables. We publish the {\it VI}
multi-epoch OGLE photometry for all objects.

We use the sample of carbon-rich long-period variables released in the
previous part of this catalog to select objects with severe drops in
luminosity, \ie with the DY-Per-like light curves. DY~Per stars are often
related to R~CrB variables. We detect at least 600 candidates for DY~Per
stars, mostly among dust enshrouded giants. We notice that our candidate
DY~Per stars form a continuity with other carbon-rich long-period
variables, so it seems that DY~Per stars do not constitute a separate group
of variable stars.}{Stars: AGB and post-AGB -- Stars: carbon -- Magellanic
Clouds}

\Section{Introduction}
R~CrB (RCB) stars are hydrogen-deficient, carbon-rich supergiants which
undergo sudden and severe declines in brightness due to the formation of
carbon dust at irregular intervals. This is a very rare type of variable
stars, with only about 50 known representatives in the Galaxy (Clayton
1996, Zaniewski \etal 2005, Tisserand \etal 2008) and 22 RCB stars known in
the Magellanic Clouds. The evolutionary status of these objects is not yet
well understood, with two proposed scenarios for the origin of the RCB
stars (\eg Iben \etal 1996): the amalgamation of a binary white dwarf
system or the expansion of a pre-white dwarf to supergiant size through the
final helium shell flash.

RCB stars can be divided into three sub-classes: hot ($\approx20\,000$~K),
warm ($\approx7000$~K) and cool ($\approx5000$~K) stars. Additionally,
there is a group of very cool objects ($\approx3500$~K), called DY~Per
stars. This last class of variables also undergoes irregular declines in
brightness, but the fading episodes are much slower than in classical RCB
stars and with symmetric recoveries. It is not clear whether DY~Per stars
are related to RCB variables, or they represent extreme cases of classical
carbon-rich (C-rich) asymptotic giant branch (AGB) stars.

The first RCB star in the Large Magellanic Cloud (LMC) -- W~Men (HV~966) --
was discovered by Luyten (1927). Before the era of large microlensing
surveys only two more variables of that type have been identified in the
LMC: HV~5637 (Hodge and Wright 1969) and HV~12842 (Payne-Gaposchkin 1971).
Both stars were confirmed spectroscopically by Feast (1972). The MACHO
microlensing project increased to 13 the number of known classical RCB
stars in the LMC (Alcock \etal 1996, 2001). They also discovered four
DY~Per stars in the LMC. The EROS-2 survey yielded additional six RCB stars
and six DY~Per variables (Tisserand \etal 2009). They also listed two
unconfirmed candidates for RCB variables and eleven candidates for DY~Per
stars in this galaxy. Thus, at present 19 ordinary RCB stars and ten DY~Per
variables in the LMC have spectroscopically confirmed status. The Optical
Gravitational Lensing Experiment (OGLE) made available a real time
monitoring system of the known RCB variables in the Magellanic Clouds and
Galactic Bulge (the RCOM system, Udalski 2008).

In this paper we describe the OGLE-III catalog of RCB stars in the
LMC. This is a part of the OGLE-III Catalog of Variable Stars (OIII-CVS)
which is intended to include all variable sources detectable in the
OGLE-III fields. In the previous part of the OIII-CVS (Soszyñski \etal
2009) we presented the sample of almost 92\,000 long-period variables
(LPVs) in the LMC, including about 9500 C-rich stars. All the confirmed and
candidate DY~Per stars described in this paper can be found in the OGLE-III
catalog of LPVs.

\Section{Observational Data}
All observations provided in this catalog were obtained with the 1.3-meter
Warsaw telescope located at Las Campanas Observatory in Chile. The
observatory is operated by the Carnegie Institution of Washington. The
telescope was equipped with the CCD mosaic camera consisting of eight
detectors with $2048\times4096$ pixels each, with the total field of view
of about $35\times35.5$~arcmin.

The third phase of the OGLE project (OGLE-III) lasted from 2001 to
2009. During this period about 40 square degrees in the LMC were
photometrically monitored for stellar variability. We collected typically
400--600 observing points per star in the {\it I} photometric band and
40--60 measurements in the {\it V}-band. For the central 4.5 square degrees
of the LMC the OGLE-III photometry has been supplemented by the OGLE-II
data collected between 1997 and 2000, which gives the total time span of
observations of over 12 years. For more information on the instrumentation
setup and the photometric reduction techniques see Udalski (2003), Udalski
\etal (2008) and the previous papers of this series.

The near- and middle-infrared single-epoch photometric measurements used in
this study were originated in the 2MASS point source catalog (Cutri \etal
2003) and the SAGE Spitzer survey (Meixner \etal 2006). The search radius
of 1\arcs\ was used to match the visual and infrared data.

\Section{Identification of Variable Stars}
\Subsection{RCB Stars}
The majority of RCB stars presented in this catalog were identified by
visual inspection of light curves during the selection of LPVs in the LMC
(Soszyñski \etal 2009). The light curves with distinct, irregular drops in
brightness were retained in a separate list and then cross-identified with
the most recent catalog of RCB and DY~Per stars in the LMC by Tisserand
\etal (2009). From 19 spectroscopically confirmed RCB stars in the LMC two
objects (HV~12842 and EROS2-LMC-RCB-6) are located outside the OGLE-III
fields, so there are no OGLE light curves for these stars. Further two RCB
stars (HV~5637, EROS2-LMC-RCB-3) do not show any significant magnitude
declines during the time span covered by the OGLE survey. The same behavior
for HV~5637 was noticed by Alcock \etal (2001) and Tisserand \etal
(2009). Even so, both stars are included in our catalog, because it is
well-known that some RCB stars may go for long intervals without minima
(\eg XX~Cam, Diethelm 1994).

\begin{figure}[p]
\centerline{\includegraphics[width=13.5cm, bb=35 55 530 745]{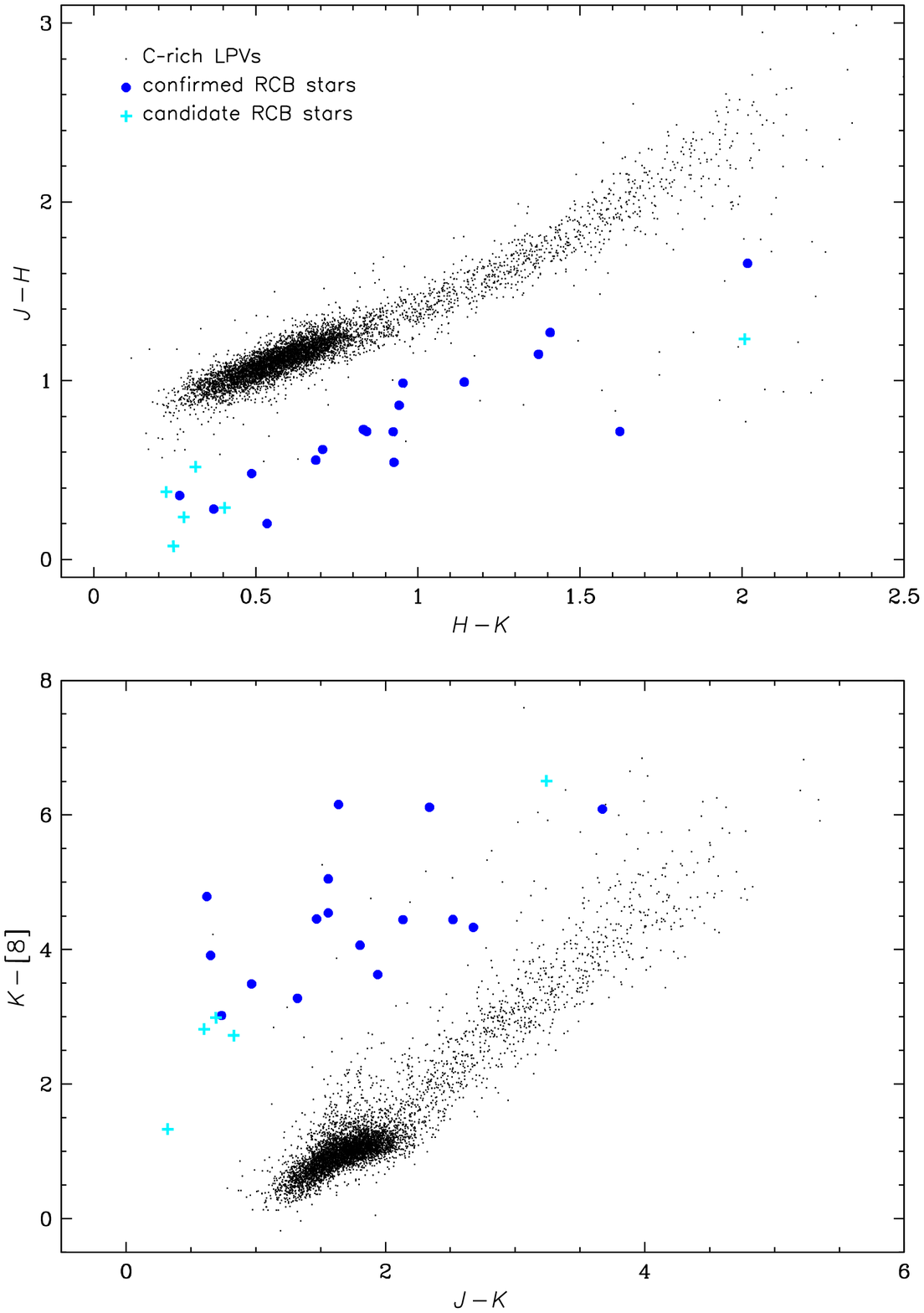}}
\FigCap{Color--color diagrams for the confirmed (blue dots) and candidate
(cyan crosses) RCB stars in the LMC. Small black dots mark C-rich AGB stars
from Soszyñski \etal (2009).}
\end{figure}
Thus, this catalog includes 17 of the 19 confirmed RCB stars in the
LMC. Their 2MASS and SAGE infrared magnitudes were used to plot the
color--color diagrams presented in Fig.~1. We also show here the C-rich
Miras and semiregular variables (SRVs) in the LMC from the previous part of
the OIII-CVS (Soszyñski \etal 2009). In these planes classical RCB stars
are well separated from the majority of AGB stars (\eg Morgan \etal 2003),
so the infrared colors can be used as a tool for selecting candidates for
RCB stars.

\begin{figure}[t]
\centerline{\includegraphics[width=11.3cm, bb=65 135 535 745]{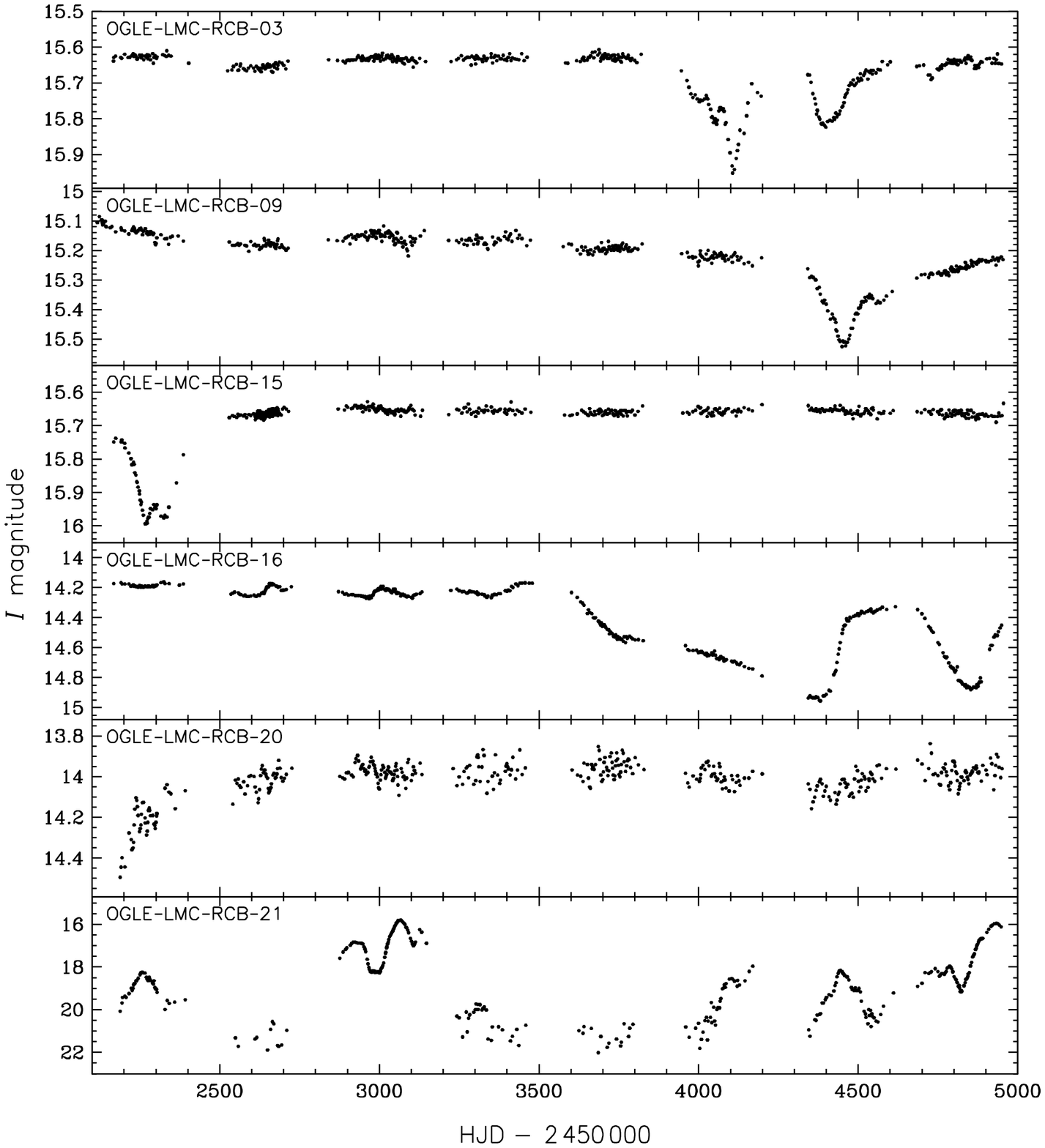}}
\FigCap{The OGLE-III {\it I}-band light curves of candidates for RCB stars
in the LMC.}
\end{figure}
In the list of irregular stars compiled on the basis of the light curve
inspection we found six other objects with the infrared colors similar to
the confirmed RCB stars. These stars are marked in Fig.~1 with cyan crosses
and their OGLE-III {\it I}-band light curves are shown in Fig.~2. As can be
seen, these stars exhibit irregular fading episodes, however usually not so
deep and not so sudden as for typical RCB stars. The only exception is
OGLE-LMC-RCB-21 with the light decline up to 5.7~mag. It looks that
virtually all {\it bona fide} RCB stars in the region covered by the
OGLE-III fields in the LMC have been already cataloged by the MACHO and
EROS-2 projects. Our six new candidates are likely low-active RCB stars,
but obviously, their status has to be confirmed spectroscopically.

\Subsection{DY~Per stars}
\begin{figure}[p]
\centerline{\includegraphics[width=13.5cm, bb=35 55 530 745]{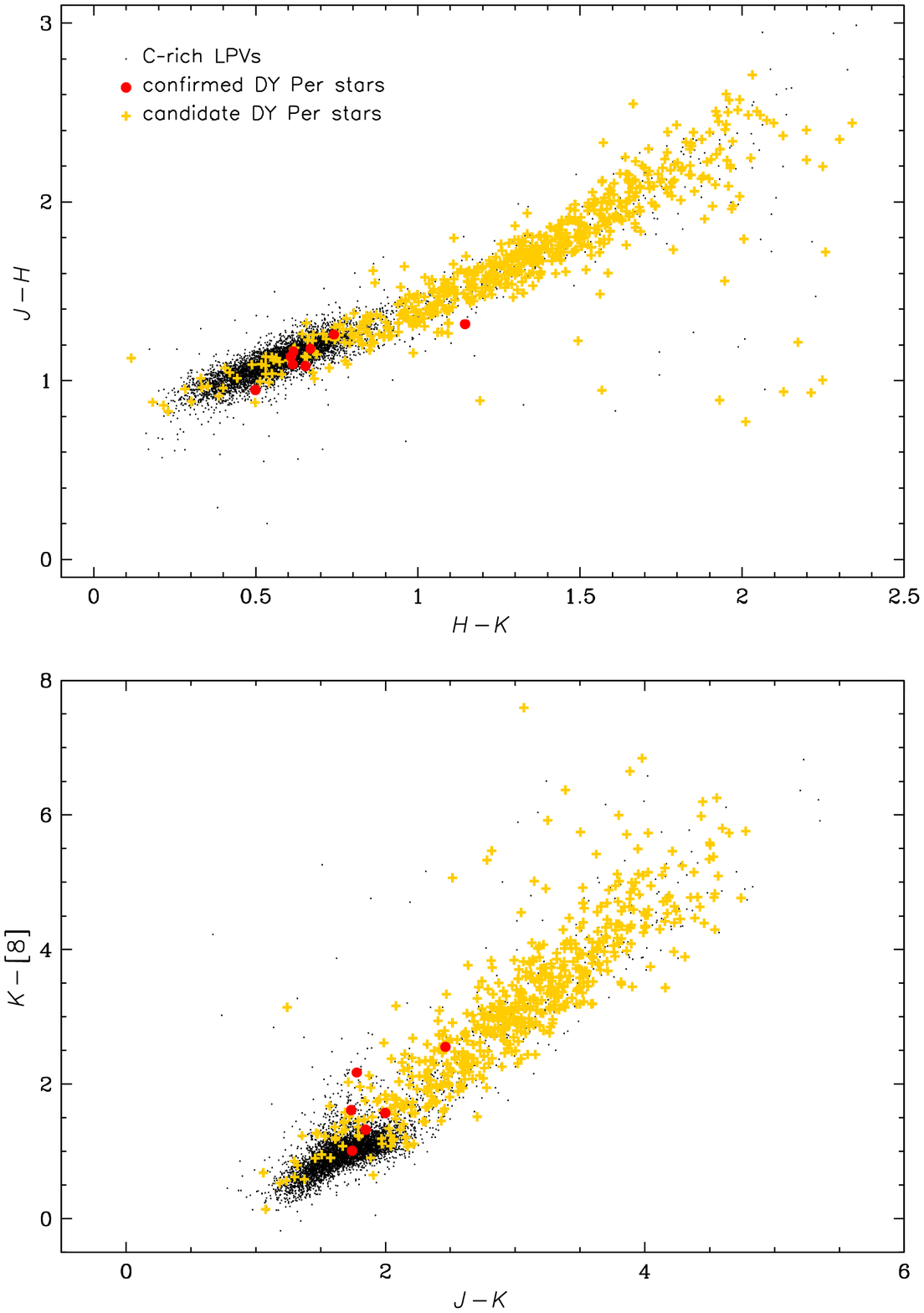}}
\FigCap{Color--color diagrams for the confirmed (red dots) and candidate
(yellow crosses) DY~Per stars in the LMC. Small black dots mark C-rich AGB
stars from Soszyñski \etal (2009).}
\end{figure}
From ten confirmed DY~Per stars (Alcock \etal 2001, Tisserand \etal 2009)
our catalog includes eight objects. EROS2-LMC-DYPer-2 and EROS2-LMC-DYPer-3
lie out of the OGLE-III fields. We marked these eight stars with red dots
in the infrared color--color diagrams (Fig.~3). As noticed by Alcock
\etal (2001) and Tisserand \etal (2009), DY~Per stars occupy the same
region in the color--color diagrams as other C-rich AGB stars. Actually,
the light curves of DY~Per stars are typical for C-rich LPVs and all of
them were included in our catalog of LPVs in the LMC (Soszyñski \etal
2009). The irregular variations of brightness superimposed on the periodic,
pulsational changes are very common among C-rich SRVs and Miras, especially
among dust-enshrouded objects.

\begin{figure}[t]
\centerline{\includegraphics[width=11.7cm, bb=65 135 535 745]{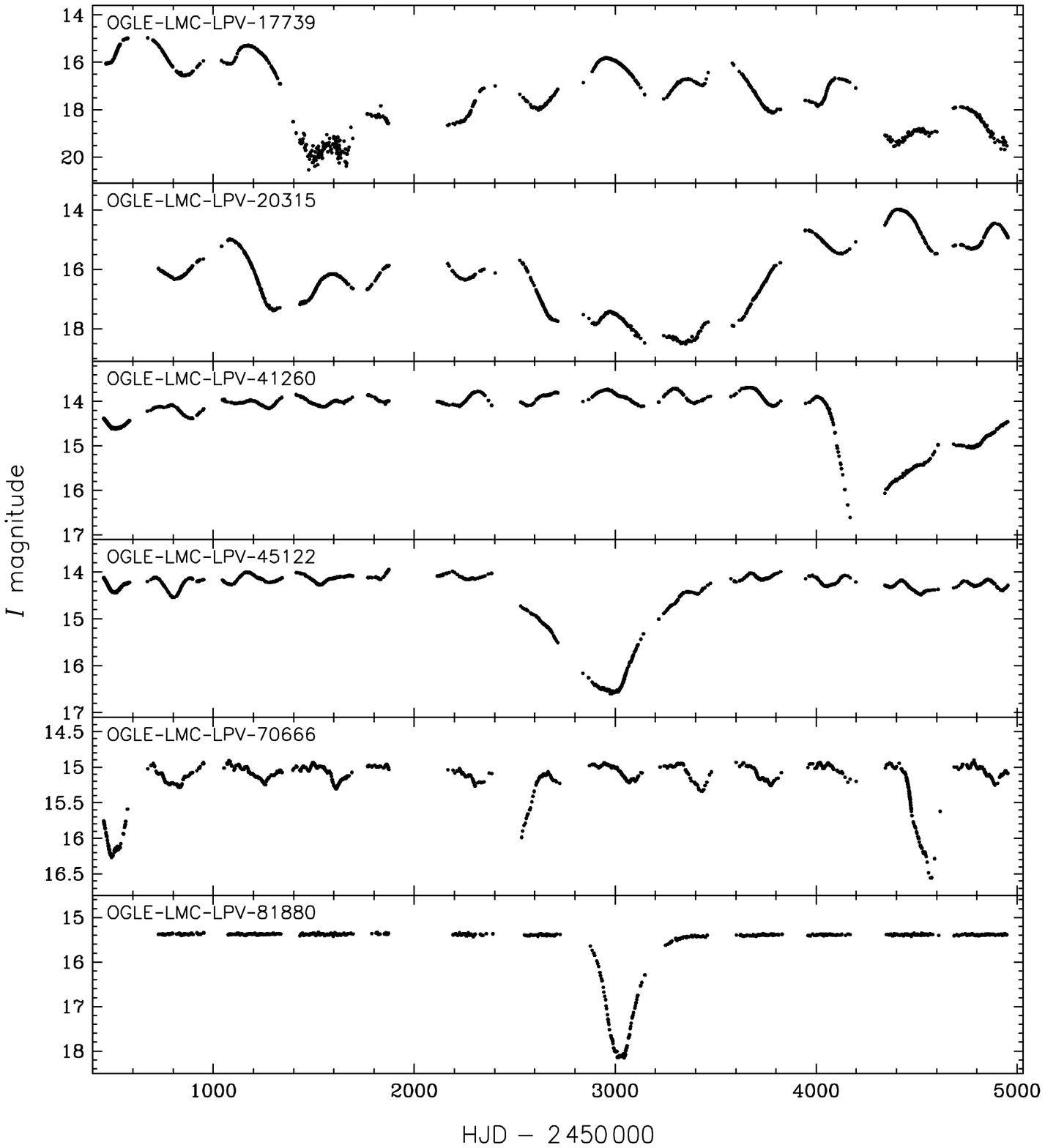}}
\FigCap{The OGLE-II and OGLE-III {\it I}-band light curves of six selected
candidates for DY~Per stars in the LMC.}
\end{figure}

We looked through the light curves of AGB LPVs in the LMC from the OGLE-
III catalog and detected 600 objects with similar features as possessed by
the confirmed DY~Per stars, \ie with significant irregular declines of
brightness ($>1.5$~mag). We marked these stars in Fig.~3 with yellow
crosses. Most of the DY-Per-like stars occupy the area of the dust obscured
AGB stars, \ie $(J-K)\apprge2$~mag and $(K-[8])\apprge1.5$~mag.

In Fig.~4 we show the OGLE-II and OGLE-III light curves of six selected
candidates for DY~Per variables. These data show a wide variety of light
curves, with various time scales and amplitudes of the fading episodes and
different behavior in the maximum brightness. We emphasize that we did not
detect any qualitative feature that separates stars with the luminosity
drops and other C-rich stars. Both groups constitute a continuity.
\vspace*{9pt}
\Section{The Catalog of RCB Stars}
\vspace*{5pt}
In Table~1 we present the full list of confirmed and candidate RCB stars
identified in the OGLE-III fields in the LMC. The stars are sorted by
increasing right ascension and designated with symbols OGLE-LMC-RCB-NN,
where NN is a two-digit consecutive number. Table~1 provides information
about the equinox J2000.0 RA/DEC coordinates of stars and their
identifications in the OGLE-III, OGLE-II, MACHO and EROS-2 databases, as
well as in the General Catalogue of Variable Stars (Artyukhina \etal
1995). In the case of OGLE-LMC-RCB-02 there is no OGLE-III identification,
because the star is too bright and saturates profile in the regular
OGLE-III photometry. However, we publish the {\sc DoPhot} profile
photometry for this object.

Table~2 contains the basic parameters of the objects in the catalog: {\it
I-} and {\it V}-band magnitudes at maximum light, pulsation periods,
amplitudes of pulsation, and the maximum drop amplitudes in the {\it
I}-band. Several very active RCB stars have not fully recovered at all
during the OGLE observations and for these objects we provide just the
maximum of measured brightness. The periods and amplitudes of pulsation are
given only for stars that have spent outside the declines enough time
to determine periods. In some cases we used the MACHO light curves to
improve the periods. Note that RCB stars usually show semiregular and
multi-periodic variations, but we provide only one, the primary period.

The content of Tables 1 and 2 can be downloaded in the electronic form
through the WWW interface or {\it via} the anonymous FTP site:
\vspace*{5pt}
\begin{center}
{\it http://ogle.astrouw.edu.pl/} \\ {\it
ftp://ftp.astrouw.edu.pl/ogle/ogle3/OIII-CVS/lmc/rcb/}\\
\end{center}
\vspace*{5pt}
The multi-epoch {\it VI} photometry and finding charts for all stars are
also available in the OGLE Internet Archive.

\begin{landscape}
\renewcommand{\arraystretch}{1.2}
\renewcommand{\TableFont}{\scriptsize}
\MakeTable{l@{\hspace{6pt}}
l@{\hspace{6pt}}
r@{\hspace{6pt}}
r@{\hspace{6pt}}
l@{\hspace{4pt}}
l@{\hspace{6pt}}
l@{\hspace{5pt}}
l@{\hspace{3pt}}
l@{\hspace{3pt}}
c@{\hspace{3pt}}
l}{12.5cm}{The identifications of RCB stars in the LMC}
{\hline
 \multicolumn{1}{c}{Star name}
&\multicolumn{2}{c}{OGLE-III ID} 
&\multicolumn{1}{c}{Status} 
&\multicolumn{1}{c}{RA} 
&\multicolumn{1}{c}{DEC} 
&\multicolumn{1}{c}{OGLE-II ID} 
& MACHO ID
&\multicolumn{1}{c}{EROS-2 ID} 
&\multicolumn{1}{c}{GCVS ID} 
&Other \\
&\multicolumn{1}{c}{Field} 
&\multicolumn{1}{c}{No} & 
&\multicolumn{1}{c}{[J2000.0]} 
&\multicolumn{1}{c}{[J2000.0]} & & & & LMC... & designation \\
\hline
OGLE-LMC-RCB-01 & LMC126.5 & 51348 & conf. & 04\uph59\upm35\zdot\ups84 &$-68\arcd24\arcm44\zdot\arcs7$ &                   &             & EROS2-LMC-RCB-3 &       & \\
OGLE-LMC-RCB-02 & LMC127.4 &       & conf. & 05\uph00\upm59\zdot\ups95 &$-69\arcd03\arcm44\zdot\arcs9$ & LMC\_SC15\_85955  & 18.3325.148 &                 & V0635 & LMV1365, HV12524 \\
OGLE-LMC-RCB-03 & LMC115.8 &   119 & cand. & 05\uph04\upm40\zdot\ups52 &$-67\arcd03\arcm11\zdot\arcs8$ &                   &             &                 &       & \\
OGLE-LMC-RCB-04 & LMC112.6 & 10912 & conf. & 05\uph10\upm28\zdot\ups52 &$-69\arcd47\arcm04\zdot\arcs4$ &                   & 5.4887.14   & EROS2-LMC-RCB-2 &       & \\
OGLE-LMC-RCB-05 & LMC109.6 & 29164 & conf. & 05\uph11\upm31\zdot\ups37 &$-67\arcd55\arcm50\zdot\arcs7$ &                   & 20.5036.12  &                 & V1394 & HV5637 \\
OGLE-LMC-RCB-06 & LMC112.2 & 21869 & conf. & 05\uph14\upm40\zdot\ups22 &$-69\arcd58\arcm39\zdot\arcs9$ &                   & 5.5489.623  & EROS2-LMC-RCB-1 &       & \\
OGLE-LMC-RCB-07 & LMC109.3 & 26874 & conf. & 05\uph14\upm46\zdot\ups44 &$-67\arcd55\arcm46\zdot\arcs7$ &                   & 16.5641.22  &                 & V1667 & LMV282, HV2379 \\
OGLE-LMC-RCB-08 & LMC100.6 & 66625 & conf. & 05\uph15\upm51\zdot\ups80 &$-69\arcd10\arcm08\zdot\arcs6$ & LMC\_SC8\_151063  & 79.5743.15  &                 &       & \\
OGLE-LMC-RCB-09 & LMC100.7 & 17279 & cand. & 05\uph16\upm48\zdot\ups11 &$-69\arcd22\arcm22\zdot\arcs0$ & LMC\_SC8\_224909  &             &                 &       & \\
OGLE-LMC-RCB-10 & LMC104.4 & 59502 & conf. & 05\uph20\upm48\zdot\ups20 &$-70\arcd12\arcm12\zdot\arcs6$ & LMC\_SC21\_243519 & 6.6575.13   &                 &       & \\
OGLE-LMC-RCB-11 & LMC104.4 & 79389 & conf. & 05\uph21\upm47\zdot\ups97 &$-70\arcd09\arcm57\zdot\arcs0$ & LMC\_SC21\_136130 & 6.6696.60   &                 & V2242 & HV942 \\
OGLE-LMC-RCB-12 & LMC161.5 & 46026 & conf. & 05\uph22\upm57\zdot\ups38 &$-68\arcd58\arcm18\zdot\arcs7$ &                   & 80.6956.207 &                 & V2314 & SHV0523154 \\
OGLE-LMC-RCB-13 & LMC164.1 & 18389 & conf. & 05\uph26\upm24\zdot\ups52 &$-71\arcd11\arcm11\zdot\arcs8$ &                   & 21.7407.7   &                 & V2643 & HV966, W Men \\
OGLE-LMC-RCB-14 & LMC161.3 & 58668 & conf. & 05\uph26\upm33\zdot\ups90 &$-69\arcd07\arcm33\zdot\arcs3$ &                   & 80.7559.28  &                 & V2629 & SHV0526537 \\
OGLE-LMC-RCB-15 & LMC160.1 & 57260 & cand. & 05\uph28\upm07\zdot\ups68 &$-68\arcd52\arcm54\zdot\arcs3$ &                   &             &                 &       & \\
OGLE-LMC-RCB-16 & LMC172.8 & 11923 & cand. & 05\uph31\upm54\zdot\ups25 &$-71\arcd53\arcm49\zdot\arcs7$ &                   &             &                 &       & \\
OGLE-LMC-RCB-17 & LMC169.7 & 62497 & conf. & 05\uph32\upm13\zdot\ups35 &$-69\arcd55\arcm57\zdot\arcs8$ & LMC\_SC2\_365505  & 81.8394.1358 &                &       & \\
OGLE-LMC-RCB-18 & LMC170.4 & 17112 & conf. & 05\uph33\upm48\zdot\ups94 &$-70\arcd13\arcm23\zdot\arcs5$ & LMC\_SC1\_206923  & 11.8632.2507 &                & V3334 & LMV535, HV2671 \\
OGLE-LMC-RCB-19 & LMC180.5 & 22569 & conf. & 05\uph39\upm36\zdot\ups99 &$-71\arcd55\arcm46\zdot\arcs8$ &                   & 27.9574.93  & EROS2-LMC-RCB-4 &       & \\
OGLE-LMC-RCB-20 & LMC178.3 &  7948 & cand. & 05\uph41\upm23\zdot\ups51 &$-70\arcd58\arcm01\zdot\arcs7$ &                   & 15.9830.5   &                 &       & \\
OGLE-LMC-RCB-21 & LMC183.5 &   614 & cand. & 05\uph42\upm21\zdot\ups92 &$-69\arcd02\arcm59\zdot\arcs3$ &                   &             &                 &       & \\
OGLE-LMC-RCB-22 & LMC185.1 & 39955 & conf. & 05\uph46\upm47\zdot\ups74 &$-70\arcd38\arcm13\zdot\arcs6$ & LMC\_SC20\_138270 & 12.10803.56 &                 &       & \\
OGLE-LMC-RCB-23 & LMC203.2 &     5 & conf. & 06\uph04\upm05\zdot\ups47 &$-72\arcd51\arcm22\zdot\arcs7$ &                   &             & EROS2-LMC-RCB-5 &       & \\
\hline}
\end{landscape}
\noindent

\renewcommand{\arraystretch}{1}
\tabcolsep=7pt
\MakeTable{lccrcc}{12.5cm}{Parameters of the RCB stars in the LMC}
{\hline
\noalign{\vskip3pt}
\multicolumn{1}{c}{Star name} & $I_{\rm max}$ & $V_{\rm max}$
& \multicolumn{1}{c}{$P_{\rm puls}$} & $A_{\rm puls}(I)$ & $A_{\rm drop}(I)$ \\
& [mag] & [mag] & \multicolumn{1}{c}{[days]} & [mag] & [mag] \\
\noalign{\vskip3pt}
\hline
\noalign{\vskip3pt}
OGLE-LMC-RCB-01 & 13.143 & 14.333 &  39.216 & 0.067 &  0.18 \\
OGLE-LMC-RCB-02 & 13.324 & 14.580 &  84.679 & 0.117 &  7.36 \\
OGLE-LMC-RCB-03 & 15.627 & 16.566 &  18.839 & 0.004 &  0.32 \\
OGLE-LMC-RCB-04 & 13.265 & 14.483 &  42.723 & 0.056 &  2.43 \\
OGLE-LMC-RCB-05 & 13.528 & 14.798 &  39.377 & 0.051 &  0.27 \\
OGLE-LMC-RCB-06 & 13.570 & 15.239 &  53.426 & 0.188 &  6.16 \\
OGLE-LMC-RCB-07 & 13.427 & 15.200 &  45.472 & 0.089 &  7.63 \\
OGLE-LMC-RCB-08 & 13.583 & 15.390 &  53.840 & 0.136 &  9.14 \\
OGLE-LMC-RCB-09 & 15.102 & 15.563 &  20.770 & 0.011 &  0.42 \\
OGLE-LMC-RCB-10 & 14.943 & 17.912 & \multicolumn{1}{c}{--} &  --   &  6.63 \\
OGLE-LMC-RCB-11 & 13.520 & 14.490 &  57.690 & 0.156 &  7.09 \\
OGLE-LMC-RCB-12 & 14.223 & 16.140 & \multicolumn{1}{c}{--} &  --   &  6.75 \\
OGLE-LMC-RCB-13 & 13.323 & 13.797 &  66.957 & 0.033 &  2.98 \\
OGLE-LMC-RCB-14 & 13.667 & 15.690 &  58.116 & 0.106 &  5.64 \\
OGLE-LMC-RCB-15 & 15.656 & 16.645 & \multicolumn{1}{c}{--} &  --   &  0.34 \\
OGLE-LMC-RCB-16 & 14.179 & 14.437 & \multicolumn{1}{c}{--} &  --   &  0.81 \\
OGLE-LMC-RCB-17 & 14.440 & 16.580 & 128.139 & 0.114 &  6.82 \\
OGLE-LMC-RCB-18 & 15.391 & 15.758 &  46.272 & 0.038 &  5.80 \\
OGLE-LMC-RCB-19 & 13.378 & 15.142 &  54.993 & 0.151 &  5.47 \\
OGLE-LMC-RCB-20 & 13.976 & 15.173 &  18.857 & 0.041 &  0.52 \\
OGLE-LMC-RCB-21 & 15.793 &   --   & \multicolumn{1}{c}{--} &  --   &  5.72 \\
OGLE-LMC-RCB-22 & 13.410 & 15.087 &  46.435 & 0.142 &  7.44 \\
OGLE-LMC-RCB-23 & 13.249 & 14.380 &  84.717 & 0.124 &  1.80 \\
\noalign{\vskip3pt}
\hline}

\begin{figure}[t]
\centerline{\includegraphics[width=12.7cm]{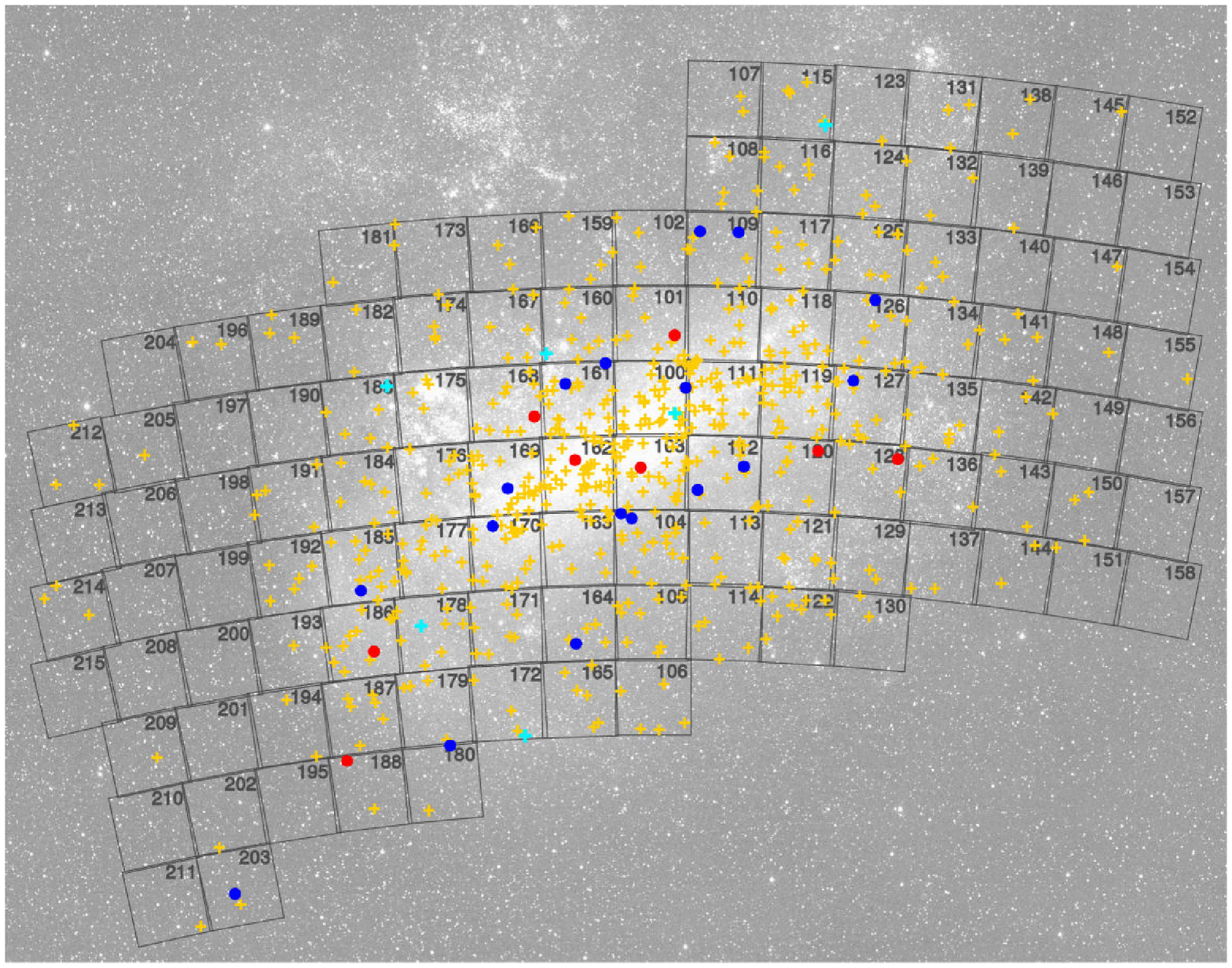}}
\vskip0.1cm
\FigCap{Spatial distribution of RCB and DY~Per stars in the LMC. Color
symbols represent the same types of stars as in Figs.~1 and~3. The
background image of the LMC is originated from the ASAS wide field sky
survey (Pojmañski 1997).}
\end{figure}
All the confirmed and candidate DY~Per stars are included in the OGLE-III
catalog of LPVs (Soszyñski \etal 2009), so we decided not to prepare the
separate list of these objects. We flagged these stars in the remarks of
the previous part of the OIII-CVS. The spatial distribution of the RCB and
DY~Per stars from our catalog is shown in Fig.~5.

\Section{Discussion}
We performed an independent search for RCB in the LMC and found no
undoubted new objects of that type. It suggests that the catalogs of RCB
stars in the LMC prepared on the basis of the MACHO and EROS-2 data are
close to being complete. The six new candidates for RCB stars identified
among the OGLE light curves have infrared colors similar to the confirmed
RCB stars, and their {\it I}-band light curves exhibit declines, however
not so deep as for typical RCB stars. We propose to consider these new
candidates as low-active RCB stars, similar to the confirmed HV~5637 and
EROS2-LMC-RCB-3. Only one candidate RCB variable -- OGLE-LMC-RCB-21 --
undergoes large brightness variations ($>5$~mag), but this star have not
reached flat maximum during the OGLE coverage, so its resemblance to RCB
stars is not firm (it may be also a DY~Per star). The status of all
candidate RCB stars must be confirmed spectroscopically.

Among these six new candidates, one object (OGLE-LMC-RCB-20 =\linebreak
KDM5651) was placed by Morgan \etal (2003) in the list of the suspected RCB
stars on the basis of its spectral characteristics. We also checked two
candidate RCB stars selected by Tisserand \etal (2009). EROS2-LMC-RCB-7
seems to be a highly-obscured star with the {\it I}-band luminosity varying
between 19~mag and 20~mag. A sign of periodicity of about 700 days and the
infrared colors of this object suggest that this is a C-rich Mira. The
other EROS candidate for RCB stars -- EROS-LMC-RCB-8 -- lies outside the
OGLE-III fields. The object MSX050825.4--685359, considered by Wood and
Cohen (2001) as a possible RCB star, seems to be a DY~Per variable.

Eight of ten DY~Per stars identified and spectroscopically confirmed by
Alcock \etal (2001) and Tisserand \etal (2009) can be detected in the LMC
OGLE-III fields. All these stars are included in the OGLE-III catalog of
LPVs in the LMC (Soszyñski \etal 2009) and classified as SRVs. Larger or
smaller declines occur in all these objects with exception of
EROS2-LMC-DYPer-1 (=OGLE-LMC-LPV-07762) which in the OGLE data shows only
small variations ($\approx0.2$~mag) of the mean luminosity.

Our photometry of C-rich AGB stars shows that severe variations of light at
irregular intervals are quite common among C-rich red giants. We selected
600 stars with DY~Per-like curves. Most of them have $(J-K)>2$~mag, \ie
they are thought to be AGB stars with thick circumstellar envelopes. We
stress that the selected objects do not constitute any separate group of
variable stars, but they form a continuity with other C-rich AGB
stars. Thus, it should be considered that DY~Per stars are just the extreme
cases of dust-enshrouded AGB stars, and they are not separate type of
variables.

The vast majority of our DY~Per candidates are Miras or SRVs, but a number
of objects were categorized as OGLE Small Amplitude Red Giants
(OSARGs). Some of these stars exhibit Long Secondary Periods (LSPs), just
like OGLE-LMC-LPV-70666 presented in Fig.~4 with the LSP equal to
$\approx1$~year. Recently, Wood and Nicholls (2009) showed that the LSP
phenomenon is associated with increased mass loss from red giants and the
circumstellar dust is likely in a clumpy or a disk-like configuration.
OGLE-LMC-LPV-70666 is a special case, because the three deep minima
recorded by OGLE seem to happen periodically, with the time interval of
about 2010~days. It is possible that we detected eclipses by a compact
cloud of dust orbiting the red giant. Another light curve which draws
particular attention is OGLE-LMC-LPV-81880 (also shown in Fig.~4). This is
likely an OSARG variable which exhibited a single, nearly symmetric dip of
almost 3~mag. It is unclear if this event was caused by a dust ejection, or
by another phenomenon.

\Acknow{We are grateful to P.~Tisserand for pointing out errors in Figs.~1
and ~3. We thank G.~Pojmañski and J.~Skowron for providing software and data
which enabled us to prepare this study.

This work has been supported by the Foundation for Polish Science through
the Homing (Powroty) Program and by MNiSW grants: NN203293533 to IS and
N20303032/4275 to AU.}

\end{document}